\documentclass[aps,twocolumn,amsmath,amssymb,pra,superscriptaddress,10pt]{revtex4-1}
\usepackage[latin1]{inputenc}

\usepackage{amssymb,amsfonts,amsmath}
\usepackage{graphicx}
\usepackage{dcolumn}
\usepackage{bm}

\newcommand{\n}{{\bf n}}

\newcommand{\ket}[1]{\ensuremath{ \ | #1 \rangle}}
\newcommand{\bra}[1]{\ensuremath{\langle #1 | \ }}

\begin{document}
\title{
Fermionic transport in a homogeneous Hubbard model:\\
 Out-of-equilibrium dynamics with ultracold atoms}
\author{Ulrich Schneider}
\email{ulrich.schneider@lmu.de}
\affiliation{Institut f\"ur Physik, Johannes Gutenberg-Universit\"at, 55099 Mainz, Germany}
\affiliation{Fakult\"at f\"ur Physik, Ludwig-Maximilians-Universit\"at, 80799 Munich, Germany}
\author{Lucia Hackerm\"uller}
\affiliation{Institut f\"ur Physik, Johannes Gutenberg-Universit\"at, 55099 Mainz, Germany}
\affiliation{School of Physics and Astronomy, University of Nottingham, NG7 2RD Nottingham, UK}
\author{Jens Philipp Ronzheimer}
\affiliation{Institut f\"ur Physik, Johannes Gutenberg-Universit\"at, 55099 Mainz, Germany}
\affiliation{Fakult\"at f\"ur Physik, Ludwig-Maximilians-Universit\"at, 80799 Munich, Germany}
\author{Sebastian Will}
\affiliation{Institut f\"ur Physik, Johannes Gutenberg-Universit\"at, 55099 Mainz, Germany}
\affiliation{Fakult\"at f\"ur Physik, Ludwig-Maximilians-Universit\"at, 80799 Munich, Germany}
\author{Simon Braun}
\affiliation{Institut f\"ur Physik, Johannes Gutenberg-Universit\"at, 55099 Mainz, Germany}
\affiliation{Fakult\"at f\"ur Physik, Ludwig-Maximilians-Universit\"at, 80799 Munich, Germany}
\author{Thorsten Best}
\affiliation{Institut f\"ur Physik, Johannes Gutenberg-Universit\"at, 55099 Mainz, Germany}
\author{Immanuel Bloch}
\affiliation{Institut f\"ur Physik, Johannes Gutenberg-Universit\"at, 55099 Mainz, Germany}
\affiliation{Fakult\"at f\"ur Physik, Ludwig-Maximilians-Universit\"at, 80799 Munich, Germany}
\affiliation{Max-Planck-Institut f\"ur Quantenoptik, 85748 Garching, Germany}
\author{Eugene Demler}
\affiliation{Department of Physics, Harvard University, Cambridge, MA 02138, USA}
\author{Stephan Mandt}
\affiliation{Institut f\"ur Theoretische Physik, Universit\"at zu K\"oln, 50937 Cologne, Germany}
\author{David Rasch}
\affiliation{Institut f\"ur Theoretische Physik, Universit\"at zu K\"oln, 50937 Cologne, Germany}
\author{Achim Rosch}
\affiliation{Institut f\"ur Theoretische Physik, Universit\"at zu K\"oln, 50937 Cologne, Germany}

\begin{abstract}Transport properties are among the defining characteristics of many important phases in condensed matter physics. In the presence of strong correlations they are difficult to predict even for model systems like the Hubbard model. In real materials they are in general obscured by additional complications including impurities, lattice defects or multi-band effects. Ultracold atoms in contrast offer the possibility to study transport and out-of-equilibrium phenomena in a clean and well-controlled environment and can therefore act as a quantum simulator for condensed matter systems.
Here we studied the expansion of an initially confined fermionic quantum gas in the lowest band of a homogeneous optical lattice. While we observe ballistic transport for non-interacting atoms, even small interactions render the expansion almost bimodal with a dramatically reduced expansion velocity. The dynamics is independent of the sign of the interaction, revealing a novel, dynamic symmetry of the Hubbard model.

\end{abstract}

\maketitle

In solid state physics, transport properties are among the key observables, the most prominent example 
being the electrical conductivity, which e.g.\ allows to distinguish normal conductors from insulators or superconductors. 
Furthermore, many of today's most intriguing solid state phenomena manifest
  themselves in transport properties, 
examples including high-temperature superconductivity, giant magnetoresistance, quantum hall physics, 
topological insulators and disorder phenomena. 
Especially in strongly correlated systems, where the interactions between the conductance electrons are important, 
transport properties are difficult to calculate beyond the linear response regime. 
In general, predicting out-of-equilibrium fermionic dynamics represents an even harder 
problem than the prediction of static properties like the nature of the ground state.
In real solids further complications arise 
due to the effects of e.g.\ impurities, lattice defects and phonons. 
These complications render an experimental investigation in a  clean and well 
controlled ultracold atom system highly desirable.
While the last years have seen dramatic progress in the control of quantum
gases in optical lattices~\cite{Jaksch, Lewenstein,Bloch:RMP}, a thorough
understanding of the dynamics in these systems is still lacking.
Genuine dynamical experiments can not only uncover new dynamic phenomena
but are also essential to gain insight into the timescales needed to
achieve equilibrium in the lattice~\cite{Hung2010,eth:doublondecay} or to
adiabatically load into the lattice~\cite{Julia,Tatjana}.

\begin{figure}[hbt]
\centering
\includegraphics[width=\columnwidth]{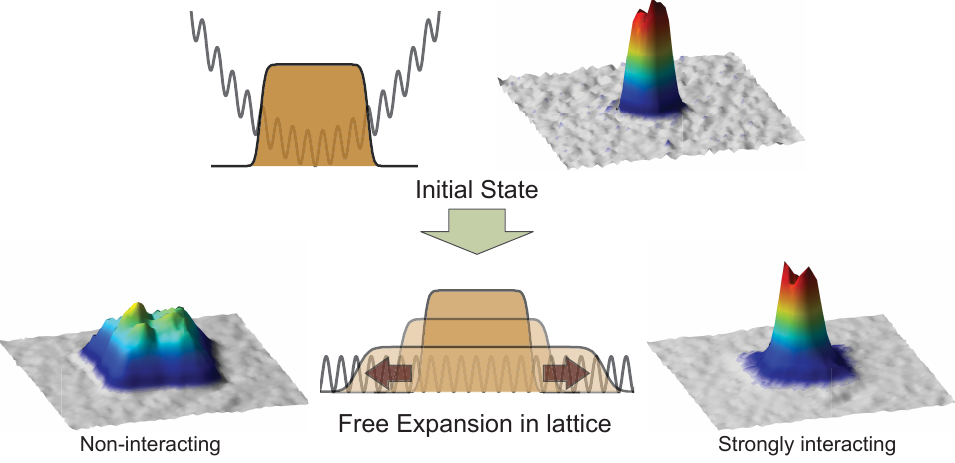}
\caption{\textbf{Expansion of fermionic atoms after a quench of the trapping potential.}
First a dephased band-insulator is created in the combination 
of an optical lattice and a strong harmonic trap. Subsequently the harmonic confinement is
  switched off and the cloud expands in a homogeneous Hubbard
  model. The observed in-situ density distributions demonstrate the 
	strong effects of interactions on the evolution.}\label{fig:intro}
\end{figure}

Using both bosonic and fermionic~\cite{Hackermueller,ETHZ:MI,schneider:MI}
atoms, it has become possible to simulate models of strongly
interacting quantum particles, for which the Hubbard model~\cite{Hubbard} is probably the most important example.
A major advantage of these systems compared to real solids is the possibility to change
all relevant parameters in real-time by e.g.\ varying laser
intensities or magnetic fields. 
While first studies of dynamical properties of both
bosonic and fermionic quantum
gases~\cite{LENS:insul,LENS:collInduced,ETHZ:transport} have already been performed,
  a remaining key challenge, however, has been the
presence of additional potentials: These will lead to confining forces 
or, in the absence of interactions, to Bloch oscillations~\cite{Lignier:dynCont,Dahan:BlochOsc,porto:osc,gustavsson:BlochOsc,fattori:BlochOsc} that dominate transport.

In this work, it was possible to study out-of-equilibrium dynamics and transport
 in a homogeneous Hubbard model by allowing an initially
confined atomic cloud with variable interactions to expand freely within a homogeneous optical
lattice (Fig.\ \ref{fig:intro}) without further potentials. 
Monitoring the in-situ density distribution during the expansion led to several surprising observations:
Already small interactions cause a drastic reduction of mass 
transport within the expanding atomic cloud and change its shape. 
For strong interactions the core of the atomic cloud does not expand, but shrinks.
And, surprisingly, we find that 
only the magnitude but not the sign of the
interaction matters: the observed dynamics is identical for repulsive and
attractive interactions despite a large difference in total energy.

The experiment starts with the preparation of a band-insulating state of fermionic potassium 
in a combination of a blue-detuned three-dimensional optical lattice
and a red-detuned dipole trap (see methods). The applied lattice loading procedure results in a cloud of atoms that are localized to single lattice sites,
where the interaction between the two used hyperfine states can be controlled using a Feshbach resonance without affecting the density distribution
(see Supplementary Information (\textit{SI} A) for details).
Subsequently, the expansion is initiated by suddenly eliminating all confining potentials in the horizontal direction (see Fig.\ 1). 
The resulting mass transport is not driven by an external potential but by density gradients.
The applied preparation scheme guarantees that all interaction effects 
arise only during the expansion since the initial state is independent of the chosen interaction.

\section{Non-Interacting case}
\begin{figure}[htb]
\centering
\includegraphics[width=\columnwidth]{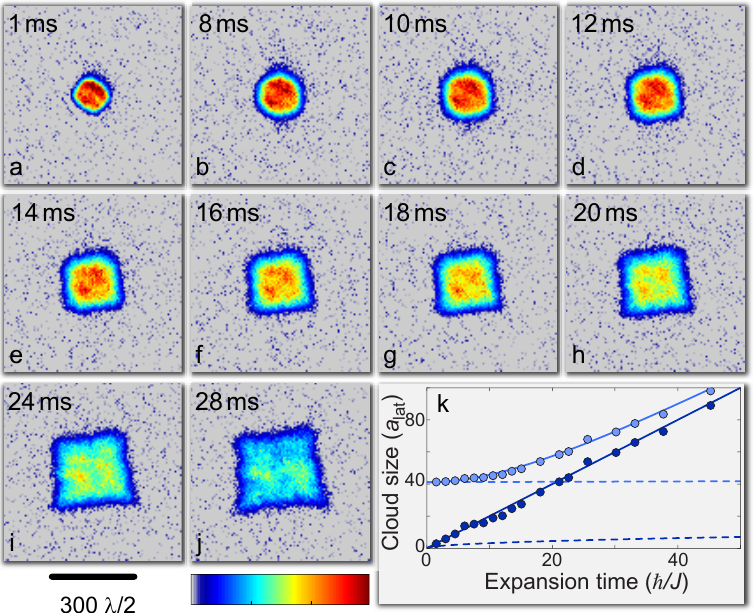}
\caption{\textbf{Expansion of Non-interacting fermions.} \textbf{a-j,} In-situ absorption images (column density a.u.) of an expanding non-interacting
  cloud in a horizontally homogeneous square lattice with lattice depth $8\,E_r$ ($1\,\text{ms}\approx1.8 \hbar/J$). The expansion changes the symmetry of
  the cloud from the rotational symmetry of the harmonic trap to the
  square symmetry of the lattice Brillouin zone. \textbf{k,} 
	Fitted cloud size $R(t)$ (light) and deconvolved single particle width 
	$R_s(t)=\sqrt{R(t)^2-R(0)^2}$ (dark) extracted from phase-contrast images. Solid lines denote the quantum mechanical prediction 
	and the dashed lines a corresponding classical random walk.}\label{fig:NonIA}
\end{figure}

For non-interacting atoms, we observe that the symmetry of the cloud
changes during the expansion from the rotational symmetry of the initial density
distribution to a square symmetry that is governed by the symmetry of
the lattice (Fig.\ \ref{fig:NonIA}). 

In the absence of collisions and additional potentials the Hubbard Hamiltonian 
consists only of the hopping term $H_J=-J\sum_{\langle i,j\rangle}\hat{c}_i^\dagger \hat{c}_j$, which describes the tunneling of a
particle from one lattice site to a neighboring site with a rate $J/\hbar$ ($\hat{c}_i^\dagger$ ($\hat{c}_i$) denotes the fermionic creation (destruction) operator).
This hamiltonian gives rise to a ballistic expansion where
each initially localized particle expands independently with a constant quasi-momentum distribution. 
Since a localized single-particle state (a Wannier function) is an equal superposition of all 
Bloch waves within the first Brillouin zone,
the velocity distribution inherits the square symmetry of the Brillouin zone.
This leads to the observed change in symmetry, as the density distribution 
after an evolution time $t$ is given by the convolution
of the initial density distribution (spherical)  with the velocity distribution (square)
 of the individual atoms 
(classically: ${\bf r}(t)={\bf r}(0)+{\bf v}t$;\hspace{3mm} ${\bf v}$: possible velocity of an individual atom,  ${\bf r}$: corresponding position).  
In the experiment, the width of a single particle wavefunction (Fig.\ 2: dark blue dots), 
which is extracted from the images by deconvolving the observed could size with the initial cloud size, 
grows linearly with expansion time, 
thereby confirming the ballistic expansion. The extracted mean expansion velocity $v_\text{exp}=\sqrt{\langle v^2 \rangle}$ agrees very well with the
quantum-mechanical prediction (solid line) $v_\text{exp}=\sqrt{2d}\,\frac{J}{\hbar}a_\text{lat}$ ($d$: dimension, $a_\text{lat}$: lattice constant),
i.e.\ the averaged group velocity of the Bloch waves (see \textit{SI} G).
This expansion can be seen as a continuous quantum walk~\cite{Aharonov,Farhi:ContQW,widera:quantumwalk,Weitenberg:addres,Childs:QWA}.
For comparison, classical (thermal) hopping of a particle 
(e.g.\ of a thermalized atom on the surface of a crystal)
would result in a random walk, where the width of the resulting density distribution 
 would scale as the square root of the expansion
time (dashed lines). For very long expansion times, residual corrugations in the
potential become relevant and can distort the square symmetry (see \textit{SI} G).

\section{Interacting case}
\begin{figure}[hbt]
\centering
\includegraphics[width=\columnwidth]{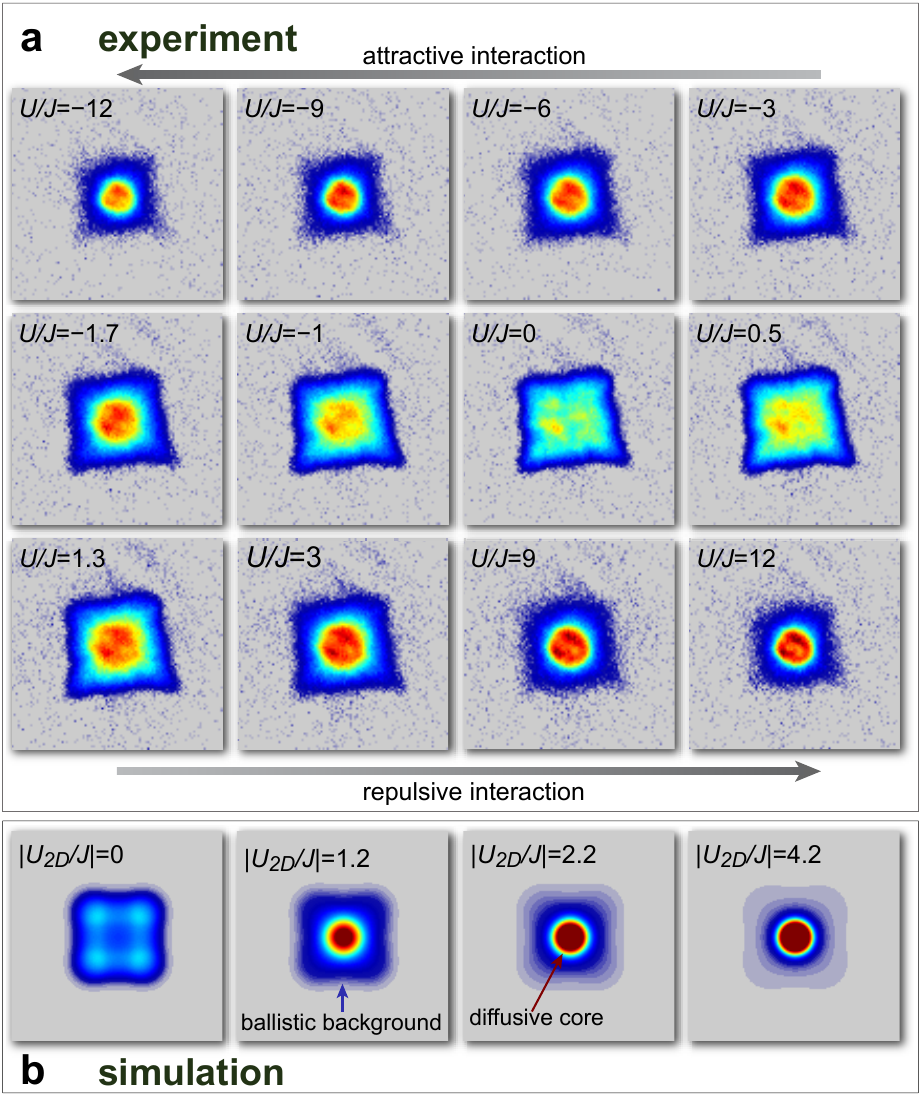}
\caption{\textbf{Expansion of interacting fermions.} \textbf{a,} In-situ absorption images for different interactions after
  $25\,$ms expansion in a horizontally homogeneous lattice.  The images show a
  symmetric crossover from a ballistic expansion for non-interacting
  clouds to an interaction dominated expansion for both attractive and
  repulsive interactions. Images are averaged over at least five
  shots and all scales are identical to Fig.\ \ref{fig:NonIA}.   
  \textbf{b,} Simulated density distributions using a 2D Boltzmann equation.}\label{fig:VarIA}
\end{figure}

The ballistic expansion observed for non-interacting atoms is in stark
contrast to the interacting case, where a qualitatively different
dynamics is observed: Fig.\ \ref{fig:VarIA} shows in-situ absorption
images taken after $25\,$ms of expansion in an $8\,E_r$ deep lattice.

The observed dynamics gradually changes from a purely ballistic expansion
in the non-interacting case into an almost bimodal expansion for interacting atoms:
Upon increasing $|U|$,  larger and larger parts of the cloud remain spherical (clearly seen in  Fig.~\ref{fig:intro})
and only a small fraction of atoms in the tails of the cloud displays a square distribution.
Here $U$ denotes the strength of the on-site interaction between different spin components ($H_{I}=U\sum_i \hat{n}_{i,\downarrow}\hat{n}_{i,\uparrow}$).
The spherical shape is a consequence of frequent collisions between the atoms in the center of the cloud, which, for the range of interactions considered here,  drive  the
system to be close to local thermal
equilibrium~\cite{Rigol:2Dtherm,werner:Therm}: 
Within the rather large clouds used in the experiment, gradients are small and the dynamics in the center can be described
by coupled non-linear diffusion equations \cite{Mandt2011}
for density $n({\bf r},t)$ and local energy $e({\bf r},t)$
\begin{eqnarray}\label{diff}
\partial_t {\bf n}=\nabla D({\bf n})\nabla {\bf n}
\end{eqnarray}
where ${\bf n}=(n,e)$ and $D({\bf n})$ is a $2 \times 2$ matrix of diffusion constants. Note that 
in the optical lattice frequent Umklapp scattering prohibits convective terms in the hydrodynamic equation (eq.\ref{diff}). Because the diffusion equation
is rotationally invariant, a diffusive dynamics can directly account for the observed spherical shape of the high density core.

For a theoretical description it is essential to realize that the diffusion equation (eq.\ref{diff}) is highly singular. Since the diffusion constant is proportional to the scattering time, it diverges with $1/n$ for small densities, $D({\bf n}) \sim 1/n$, as the probability to scatter from other atoms  is linear in $n$ for small densities. Such highly singular ``superfast'' diffusion equations have been extensively studied in the mathematical literature~\cite{vasquez06}. Remarkably, they predict a completely unphysical behavior in large dimensions ($d\ge 2$): The particle number is not conserved as particles vanish at infinity with a constant rate (for $d=2$). Due to this breakdown of the hydrodynamic approach, the expansion is not governed
by the diffusion equation but instead by the physics in the tails of the cloud where no local equilibrium can be reached.
In this regime, the densities are low and atoms scatter so rarely that their motion again becomes ballistic. Therefore the tails of the cloud show the square symmetry characteristic for freely expanding particles (Fig.\ \ref{fig:VarIA}). 
This initial fraction of ballistically expanding atoms decreases 
for increasing interaction strengths. During the
expansion the density gets reduced and, in the limit of 
infinite expansion times, all atoms are expected to become ballistic. 
This crossover into ballistic behavior for small densities leads to a breakdown of the
diffusive behavior and regularizes the otherwise singular diffusion equation.

To describe both the diffusive and the ballistic regime, we use numerical simulations based on the
semiclassical Boltzmann equation in the relaxation time approximation:
\begin{align} \label{boltzmann}
\partial_t f_{\bf {q}}+{\bf v_q} \nabla_{\bf r} f_{\bf {q}}+{\bf F}({\bf r}) \nabla_{\bf q} f_{\bf {q}}=-
\frac{1}{\tau(\n)} \left(f_{\bf {q}}-f^0_{\bf {q}}(\n) \right)
\end{align}
This equation  describes the evolution of a semi-classical 
momentum distribution $f_{\bf q}({\bf r},t)$ as a function of position 
and time in the presence of a force ${\bf F}$. 
Here the transport scattering time $1/\tau({\bf n})$, which describes the relaxation 
towards an equilibrium Fermi distribution $f^0_{\bf q}$ for given energy and particle 
densities
is determined  from a microscopic calculation of the
diffusion constant for small interactions  (see \textit{SI} B for details). 
The Boltzmann equation describes qualitatively and semi-quantitatively the observed cloud shapes, see Fig.~\ref{fig:VarIA}b.

\begin{figure}[hbt]
\centering
\includegraphics[width=\columnwidth]{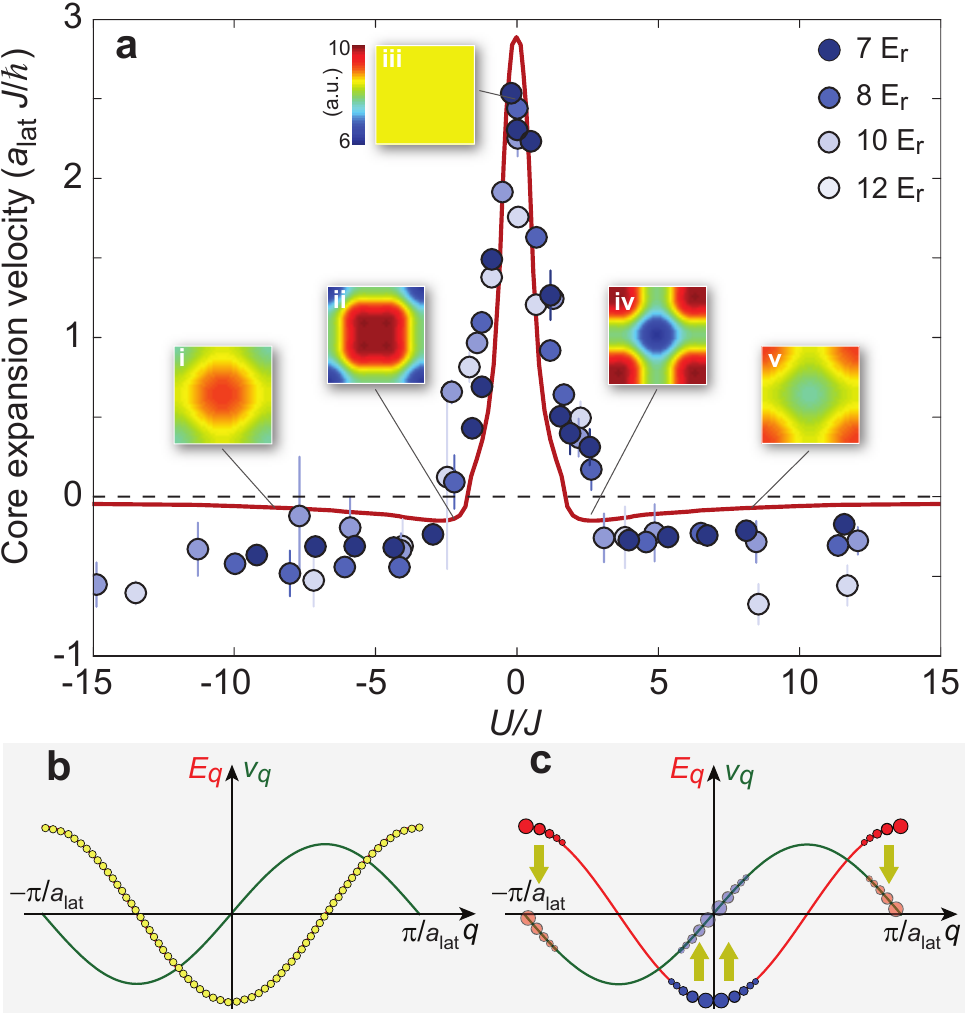}
\caption{\textbf{Core expansion velocities.} \textbf{a,} Measured core expansion velocities versus interaction for 
various lattice depths in a 2D situation (vertical lattice depth $20E_r$). 
The red line denotes the result of a numerical calculation 
(see text). 
The insets i-v show the numerically calculated quasi-momentum distribution 
after 40 ms of expansion for $U/J=-8,-2,0,2,8$. \textbf{b-c,} 1D 
energy dispersion  (red line) and group velocity (green line) 
together with schematic sketches of the relative occupations 
(yellow: initial state and $U=0$, red: $U>0$, blue: $U<0$):
While the quasi-momentum distribution is different for $U$ and $-U$, 
the resulting group velocity distribution (shaded) is identical.}
\label{fig:slopeIA}
\end{figure}

The core width $R_{c}(t)$, which measures the size of only the high
density core, is extracted from phase-contrast images by determining
 the half width at half maximum (HWHM) of the density distribution (see \textit{SI} E). 
By fitting the evolving core width to $R_c(t)=\sqrt{R_{c,0}^2+v_c^2 t^2}$, we extract the
 core expansion velocities $v_c$, which are shown in Fig.\ \ref{fig:slopeIA}.
Surprisingly, they decrease dramatically already for interactions much smaller
than the bandwidth $8 J$, which highlights the strong impact of moderate interactions on
mass transport in these systems. 
We observe the same behavior irrespective of the sign of the interactions.

For interactions larger than $|U/J|\gtrsim 3$, the dynamics of the high density core changes qualitatively: The core starts shrinking instead of expanding and the core expansion velocities $v_c$ become negative.
In this regime, the expansion of the diffusive core is strongly suppressed and the essentially 
frozen core dissolves by emitting ballistic particles and therefore shrinks in size, similarly to a melting ball of ice. This feature is also recovered by our simulations based on the Boltzmann equation (red line in Fig.\ \ref{fig:slopeIA}).
The slight asymmetry at large interactions can
be attributed to interaction dependent losses caused by
light-assisted collisions during the preparation sequence.

This pronounced dependence of the dynamics on small interactions enabled us to measure the
zero crossing of the scattering length
$(B(a=0)=209.1\pm0.2\,\text{G})$, which corresponds to a width of the Feshbach resonance of
$w=7.0\pm0.2\,\text{G}$, see \textit{SI} H.
In contrast to the high interaction limit, where the exponentially long lifetime~\cite{eth:doublondecay} 
of excess doublons leads to two independent dynamics of doublons and single atoms, 
we observe thermal equilibrium between doublons and unpaired atoms, as shown in detail in the Supplementary Information (\textit{SI} I).

We have shown that the observed transport properties can be qualitatively predicted by the semiclassical Boltzmann equation (eq.\ref{boltzmann}). 
However, the full quantum dynamics is certainly more
complex and includes e.g.\ the formation of entanglement between distant
atoms~\cite{Sanpera:ent} as well as the existence of bound or repulsively bound states.
While the expansion can be modeled in 1D~\cite{Kajala2011} using DMRG methods~\cite{Schollwoeck2005},
so far no methods are available to calculate the dynamics quantum-mechanically in higher dimensions.
The separation between ballistically expanding atoms carrying high entropy and the high density core in the center could be used to locally cool the atoms via quantum distillation processes~\cite{Fabian:QDist}.

Surprisingly, we observe identical density profiles and expansion
rates for repulsive and attractive interactions of the same strength
(see Fig.\ \ref{fig:VarIA}, \ref{fig:slopeIA}). While scattering
  cross sections are  proportional to $U^2$ for small $U$, the
  interaction energy and density gradients give rise to forces
  linear in $U$: Repulsive interactions create a
  positive pressure, which in free space would lead to an increased
  expansion rate, while an attractive interaction is expected to slow down the
  expansion, in contrast to the observed behavior in the lattice.

  The identical evolution of the density for positive and negative $U$
  is the consequence of an exact dynamical symmetry of the Hubbard
  model, which relies on two facts: First, if both the initial
  state and the observable are invariant under time reversal symmetry, 
	the dynamics of the observable is necessarily
  unchanged by the transformation $H \to -H$. 
	Second, as $E({\bf q})=-2 J \sum_i \cos(q_i d)$ is
  the kinetic energy of the Hubbard model, the sign of the
  hopping $J$ can be changed, $J \to -J$, by shifting all momenta
  $\mathbf{q} \rightarrow \mathbf{q}+(\pi,\pi,\pi)/d$.  
	If now both the initial state
  and the observable are invariant under both time reversal and the above shift of momenta, one
  necessarily obtains the same evolution for $U$ and $-U$. 
	This is the case in our experiment, initially all particles are
  localized and all momenta are therefore equally occupied while the density operator is time and momentum independent.
	A formal proof of this argument is given in the Supplementary Information (\textit{SI} D).

  During the expansion, when the density of
  atoms is reduced, interaction energy is converted into kinetic
  energy. Initially the kinetic energy of the localized particles is
  zero. For $U<0$, the total energy is therefore negative and low
  momentum states become more populated during the expansion. For
  $U>0$, in contrast, the total energy is positive, implying an
  enhanced occupation of higher momentum states. More precisely, the
  two momentum distributions for $U$ and $-U$ are  shifted by
  $(\pi,\pi,\pi)/d$, as can be seen in the insets i-v in Fig.\ \ref{fig:slopeIA}.  
	In free space, where $E({\bf q})\sim q^2$,
  larger momenta imply larger group velocities ${\bf v}({\bf q})=(1/\hbar)\,d
  E/d{\bf q}\sim q$ and the cloud expands faster for
  repulsive interactions.  For the Hubbard model, in contrast,
  the group velocities for ${\bf q}$ and $(\pi,\pi,\pi)/d-{\bf q}$ are
  the same, $v_i\sim\sin{q_i d}$, leading to the same expansion of the cloud for $U$ and $-U$, see
  Fig.~\ref{fig:slopeIA}b and c.

 Since large parts of the cloud are expected to be in local equilibrium
in the interacting case, one can define local temperatures:
For $U<0$ the
system cools down while expanding and positive local temperatures
  $0<T(r)<\infty$ are obtained. For $U>0$, in contrast, the exact dynamical
  symmetry implies that the local temperatures have to be negative as
  $\exp[-H/k_B T]=\exp[-(-H)/(-k_B T)]$. This has also been confirmed
  by our numerical calculations (see Fig.\ 7 in \textit{SI}). Negative temperatures describe equilibrated
  systems with population inversion and are well defined for systems
  like the Hubbard model where the energy has an upper bound~\cite{Rapp2010}. They have
  been observed in spin systems~\cite{purcell51} and localized ultracold atoms~\cite{Medley:11}. Assuming local thermalization, the observed $U\leftrightarrow-U$
  symmetry directly implies negative temperatures for repulsive
  interactions at long expansion times. 

\section{Conclusion}
Ultracold fermions in optical lattices offer many unique possibilities to
study non-equilibrium dynamics as they allow for a full real-time
control of almost all relevant parameters, including
quantum quenches, where the Hamiltonian of the system is
changed instantaneously. We studied the expansion of a cloud of
initially localized atoms in a homogeneous Hubbard model following a 
quench of the trapping potential and observed
the crossover from a ballistic expansion at small densities or
vanishing interactions to a bimodal expansion in the interacting
case. 
We observed identical behavior for both attractive and
repulsive interactions, highlighting the high symmetry of the
kinetic energy in the Hubbard model.
The surprisingly large observed timescales of mass transport 
set lower limits on the timescales needed both to
adiabatically load the atoms into the lattice and to
cool the system in the lattice~\cite{McKay2010} and are therefore
of paramount importance for all attempts to create complex, strongly
correlated many-body states like N\'{e}el-ordered states in these systems.

The method of directly measuring the expansion dynamics can be used
to detect complex quantum states including Mott-insulating states~\cite{schneider:MI} 
or to possibly distinguish pseudogap~\cite{Hackermueller} from superfluid states
in the attractive Hubbard model. 
In addition, the effects of various disorder potentials on the 
two-dimensional dynamics can be studied.
The extension to a Bose-Fermi mixture could enable studies on ohmic transport,
where the bosons assume the role of the phonons.

\section{Methods}
The Supplementary Information \textit{SI} contains a detailed description of the experimental sequence (A), 
the derivation of the parameters of the Boltzmann equation including the transport scattering rates (B),
simulated temperature distributions during the expansion (C), a formal proof of the observed dynamical symmetry (D),
and further experimental details regarding image processing (E), expansion velocities of non-interacting atoms (F), canceling the harmonic confinement (G), the width of the Feshbach resonance (H) and the observed doublon dissolution time (I).

\paragraph{Experimental sequence}
We use a balanced spin mixture of the two lowest hyperfine states 
$|\smash F,m_F\rangle=|9/2,-9/2\rangle$ and $|9/2,-7/2\rangle$ of fermionic
potassium $^{40}\text{K}$ with $N=2\smash-\smash3\times10^5$ atoms 
at an initial temperature of $T/T_F=0.13(2)$.
Starting in an harmonic trap, the atoms are loaded into a combination of a blue-detuned 
three-dimensional optical lattice with lattice constant $a_\text{lat}=\lambda/2=369\,\text{nm}$ 
and a red-detuned dipole trap, using a sequence similar to the one applied in~\cite{schneider:MI}.
Once in the lattice, tunneling is strongly reduced by increasing the lattice depth to $20\,E_r$ (recoil energy 
$E_r=h^2/(2m\lambda^2)$) and the interactions
 can be controlled via a Feshbach resonance without affecting the density distribution (see \textit{SI} A).
In order to initiate the expansion, the lattice depth is is lowered again and the harmonic confinement (see Fig.\ 1a) along the horizontal directions is eliminated
 by reducing the strength of the dipole trap by more than
90\%, such that along the horizontal directions the remaining dipole 
potential precisely compensates the anticonfinement produced by the 
lattice beams (see \textit{SI} G). 

While any vertical motion is expected to 
be strongly suppressed by gravity-induced Bloch oscillations (oscillation amplitude $2J/(mg)<2\,d$), 
the atoms are exposed to a homogeneous Hubbard model without additional potentials
in the horizontal directions. The evolution of the density distribution during the 
following expansion was monitored by 
in-situ imaging along the vertical axis of the cloud,
thereby integrating over any vertical dynamics. 
Absorption images of the resulting dynamics are shown in Fig.\ \ref{fig:NonIA} for
the case of non-interacting particles and in Fig.\ \ref{fig:VarIA} for
various interactions. 

For a quantitative analysis (Fig.\ 2k and 4) (see \textit{SI} E\&F), vertical tunneling of the atoms during the expansion was 
additionally suppressed
by increasing the depth of the vertical lattice to $20\,E_r$, thereby realizing several layers of independent 
two-dimensional Hubbard models. All quantitative analysis were performed using phase-contrast images.

\section{Supplementary Information}

\subsection{Experimental sequence}\label{experimentalSequence}
The experiment starts with the preparation of a band-insulating state
of a balanced spin mixture of fermionic potassium $^{40}\text{K}$, as described
in previous work~\cite{schneider:MI}: Through evaporative cooling in a
red-detuned crossed dipole trap a quantum degenerate mixture of the
two lowest hyperfine states of potassium was reached with atom numbers
of $N=1\smash-\smash1.5\times10^5$ atoms per spin state at a
temperature of $T/T_F=0.13(2)$, where $T_F$ denotes the Fermi
temperature in the harmonic trap. The trapping frequencies of the
dipole trap were then increased to $2\pi\times100\,$Hz
($2\pi\times400\,$Hz) in the horizontal (vertical) directions.

\begin{figure}[hbt]
\centering
\includegraphics[width=.9\columnwidth]{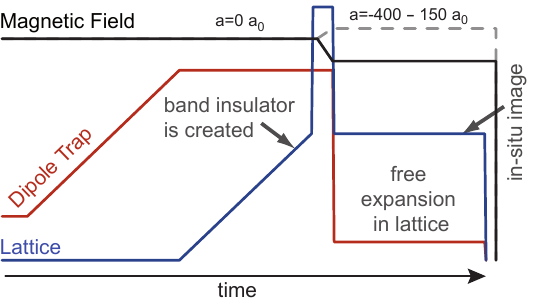}
\caption{Experimental Sequence. Starting with a degenerate
  Fermi gas in the dipole trap a non-interacting band insulator is
  created. During a freeze-out period the atoms localize to individual
  lattice sites and the desired interaction is set without altering
  the density distribution. Subsequently the harmonic confinement is
  switched off and the cloud expands in a homogeneous Hubbard
  model.}
\end{figure}

Subsequently, a simple-cubic blue-detuned 3D optical lattice
($\lambda=738\,$nm) is ramped up linearly to a depth of $8E_r$
$(1E_r=h^2/(2m\lambda^2))$ in $56\,$ms while the magnetic field is held
at $209.1\,\text{G}$, which corresponds to vanishing interactions (see~\ref{sec:FR}). 
This loading procedure results in a band-insulating state
surrounded by a thin metallic shell at a compression
of $E_t/12J=1.8$, using units and conventions of~\cite{schneider:MI}.  In the next step, the tunneling
rate $J$ is reduced to $J=h\times23\,$Hz by linearly increasing the lattice
depth to $20\,E_r$ in $200\,\mu$s, a timescale that is slow enough to
avoid excitations into excited bands, but fast compared to tunneling
within the lowest band. Due to this reduced tunneling rate the density
distribution is essentially frozen out during the following $40\,\text{ms}$
magnetic field ramp ($B_{fin}=206-260\,$G), which sets the interaction
for the expansion due to the well known Feshbach resonance at $B_0=202.1\,\text{G}$
~\cite{JILA:FR}. Combined with the strong harmonic confinement, this
leads to a dephasing between different
lattice sites and effectively localizes all particles to individual sites~\cite{will:multiorb}.

In total, this sequence produces a cloud of localized atoms with a
well-known density distribution that is independent of the interaction
between the particles. 

The expansion is initiated by lowering the
lattice depth in $200\mu s$ to values between $4E_r$ and $15E_r$ while
simultaneously switching off the harmonic confinement.  To this means,
the strength of the dipole trap is reduced by more than 90\%, such
that the remaining potential precisely compensates the anticonfinement
produced by the lattice beams (see Sec.~\ref{sec:cancelHarm}).

\subsection{Boltzmann equation} 
For a theoretical description of the expanding clouds one needs an
approach that can correctly describe both the diffusive
and the ballistic regime.
The probably simplest one is the Boltzmann Equation which we employ together with
a version of the relaxation time approximation that conserves both energy and momentum:
It is constructed to be exact both in the
diffusive and the ballistic regime and is able to provide a decent
qualitative (but not quantitative) description of the crossover
regime.
 Boltzmann
equations \cite{ziman} describe the evolution of the quasiclassical
momentum distribution $f_{\bf {q}}({\bf r},t)$ as a function of
position and time:
\begin{align} \label{boltzmannAP}
\partial_t f_{\bf {q}}+{\bf v_q} \nabla_{\bf r} f_{\bf {q}}+{\bf F}({\bf r}) \nabla_{\bf q} f_{\bf {q}}=-
\frac{1}{\tau(\n)} \left(f_{\bf {q}}-f^0_{\bf {q}}(\n) \right)
\end{align}
Here ${\bf v_q}=\nabla_{\bf q} \epsilon_{\bf q}$ denotes the velocity of
the particles and ${\bf F}({\bf r})$ describes a force term arising from the
interaction of the atoms. For the initial condition we take into
account that coherences between neighboring sites are efficiently destroyed 
during the preparation of the initial state, as
described in Sec.~\ref{experimentalSequence}. 
Therefore we use $f_{\bf {q}}({\bf r},t=0)=n({\bf r})$ 
where $n({\bf r})$ is the density distribution
of non-interacting fermions with an entropy per particle of $1.141\,k_B$. 
This entropy corresponds to an initial temperature of $T/T_F\approx 0.125$ in the harmonic trap,
which is compatible with the experiment. In the force term, we
use the Hartree approximation, ${\bf{F}}=-U \nabla n$, where $n$ is the
density per spin, but our numerical results show that this term is relatively small.  
In order to obtain the correct hydrodynamics,
the collision term on the right-hand side of Eq.~\ref{boltzmannAP}
needs to conserve both particle number and energy, and to correctly 
describe the relaxation towards equilibrium. This is achieved by setting 
\begin{equation}
f^0_{\bf
  {q}}(\n)=1/(\exp[(\epsilon_{\bf q}-\mu({\bf r},t))/k_B T({\bf r},t)]+1)
  \end{equation}
where $\mu({\bf r},t)$ and $T({\bf r},t)$ are chosen such that both
$\sum_q f_{\bf {q}}-f^0_{\bf {q}}(\n)=0$ and $\sum_q \epsilon_{\bf q}
\left(f_{\bf {q}}-f^0_{\bf {q}}(\n) \right)=0$. The effective
temperatures $T({\bf r},t)$ obtained from this procedure are, for our
initial conditions, at least an order of magnitude larger than the
bandwidth. This implies that energy conservation and the diffusion of
energy only play a minor role in the experiment, as we checked numerically.

The effective scattering rate $1/\tau(\n)$ should be chosen such that
the correct diffusion constant $D(\n)$ is obtained.  Unfortunately, we
are not aware of any numerical or analytical method to calculate
$D(\n)=\tau(\n) \langle {\bf v_q}^2 \rangle/d$ for the two-dimensional
Hubbard model, even in the limit $T \to \infty$, where all thermodynamic
properties are exactly known. Note that, for example, dynamical mean
field theory, which successfully describes thermodynamic properties
\cite{schneider:MI}, should not be used to
calculate transport properties quantitatively, as it neglects vertex
corrections that are quantitatively important even for $T\to \infty$.

 We therefore calculate the diffusion constant perturbatively to
 leading order in the interaction.  This can be done by computing
 the conductivity $\sigma$ from the translationally invariant Boltzmann
 equation~\cite{ziman}, using the full collision term with golden-rule
 transition rates. Subsequently, one can obtain $D(\n)$ and $\tau(\n)$
 from the Einstein relation $D=\sigma \partial
 \mu/\partial n$.
 \begin{figure}
\centering
\includegraphics[width=.9 \columnwidth]{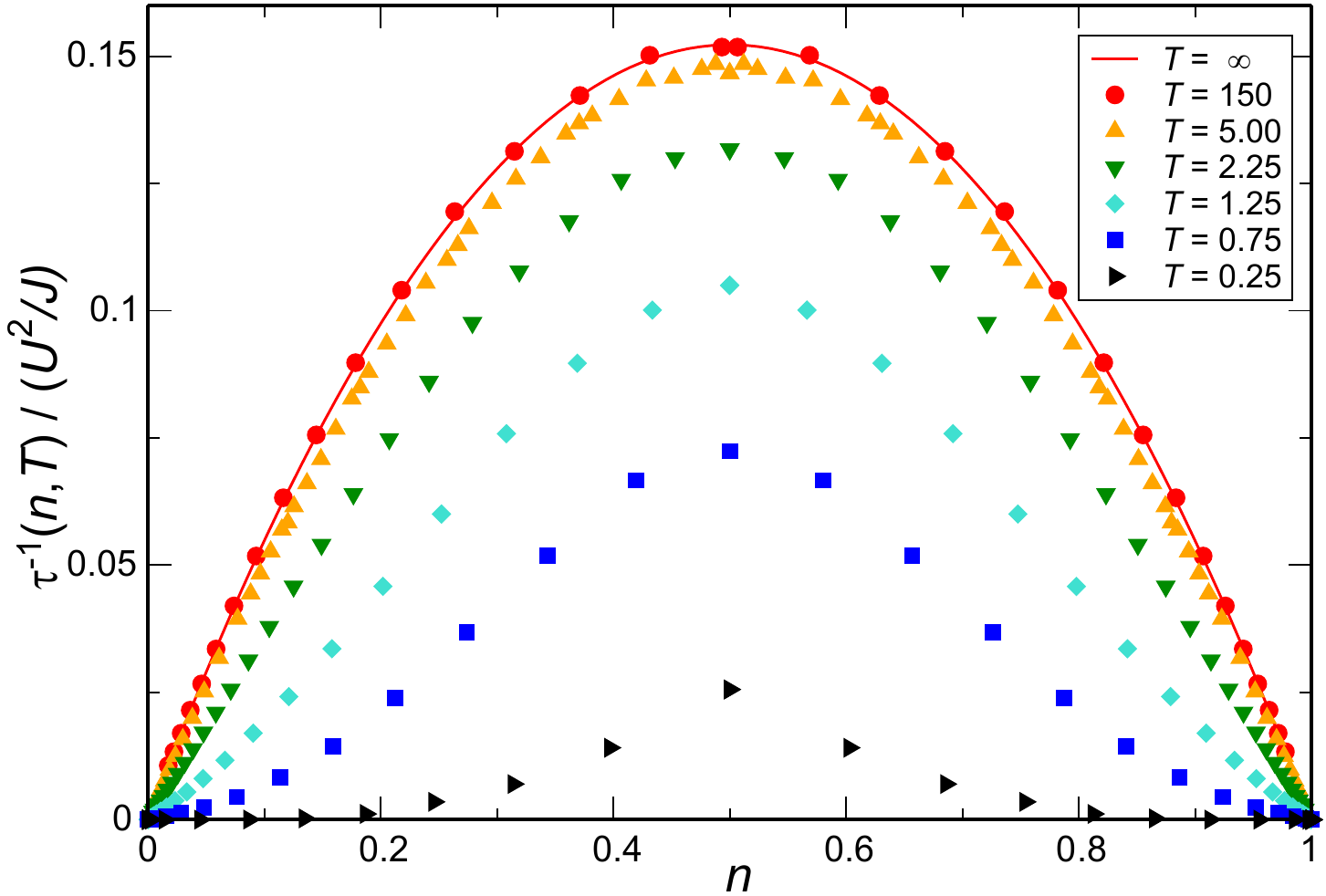}
\caption{Transport scattering rates as a function of density for
  different temperatures. Note that the transport scattering rate (in
  contrast to the single-particle scattering rate) becomes
  exponentially small in the low-density, low-temperature regime where
Umklapp scattering is suppressed.}
\label{fig:Scatt}
\end{figure}

In order to avoid the solution of a complicated integral equation, we employ
 a variational solution of the Boltzmann equation (see
\cite{ziman} for details) using $f_{\bf {q}}=f^0_{\bf {q}}-\partial
f^0_{\bf {q}}/\partial \epsilon_{\bf q} \sum_i \alpha_i c^{(i)}_{\bf
  {q}}$ with four variational parameters $\alpha_i$, and $c^{(1)}_{\bf
  {q}}=v^x_{\bf {q}}$, $c^{(2)}_{\bf {q}}= \epsilon_{\bf {q}} v^x_{\bf
  {q}}$, $c^{(3)}_{\bf {q}}=q_x$, $c^{(4)}_{\bf {q}}=(\pi/a-q_x) \
{\rm mod} \ 2 \pi/a$. Here $c^{(1)}$ and $c^{(2)}$ are chosen to enable the
 calculation of both thermal and charge diffusion constant as well as the
cross terms. The remaining terms, $c^{(3)}$ and $c^{(4)}$, on the other hand, are essential to 
correctly describe the low temperature limit: For a low density of particles or holes, 
the conductivity, and therefore the diffusion constant, grows exponentially for low
$T$ due to an exponential suppression of Umklapp scattering processes~\cite{ziman}. 
Since we determine $1/\tau(\n)$ such that we
recover the correct diffusion constant, this effect is fully included. 
Note that it is not captured by a more conventional version
of the relaxation time approximation~\cite{ziman}, which neglects
vertex correction and identifies $1/\tau(\n)$ with a single-particle
relaxation rate instead of a transport scattering rate.
Although the variational approach only gives a lower bound for the
diffusion constant, we expect it to be accurate within
a few percent for small interactions, as we have checked by reducing
 the number of variational parameters. The density dependence
of the resulting scattering rate is shown in
Fig.\ \ref{fig:Scatt} for various temperatures. 
For $E=0$ or, equivalently, $T\to \infty$ 
we obtain $1/\tau(n,E=0)\approx 0.609\, n (1-n) U^2/J$.  
To solve the resulting Boltzmann
equation (Eq.~\ref{boltzmannAP}), a simple Runge-Kutta integration is used,
discretizing both the position and momentum variable. 

For all quantitative comparisons with the experimental results we also take
into account the necessity to average over a stack of independent
two-dimensional systems with different atom numbers. This averaging, however,
changes the result only slightly since the main contributions arise from
the central layers.

All qualitative features seen in the experiment are well
reproduced by the numerical results using the Boltzmann equation. 
Fig.\ 4 in the main text shows that the numerical simulations describe
the drastic collapse of the expansion velocities and
the shrinking of the core width for strong interactions also semi-quantitatively.  
Quantitative discrepancies between experiment and numerics
 probably arise both because the leading order perturbation theory 
is not valid for $U\gtrsim J$ and because the relaxation time approximation breaks 
down in the crossover region from diffusive to ballistic behavior, where
 the colliding atoms are far from thermal equilibrium.

\subsection{Negative Temperatures}
For finite interaction strength and densities, the frequent collisions between the atoms
ensure them to stay close to local thermal equilibrium and allow for the definition of a
local temperature $T({\bf r},t)$, which, in contrast to a system in global equilibrium, 
will depend on the position $\bf r$ within the cloud.
In the Boltzmann calculation (see above), $T({\bf r},t)$ is given by the temperature of a homogeneous 
reference ensemble in equilibrium 
at the same particle ($n({\bf r},t)$) and energy density ($e({ \bf r},t)$).

\begin{figure}[htb]
\centering
\includegraphics[width=\columnwidth]{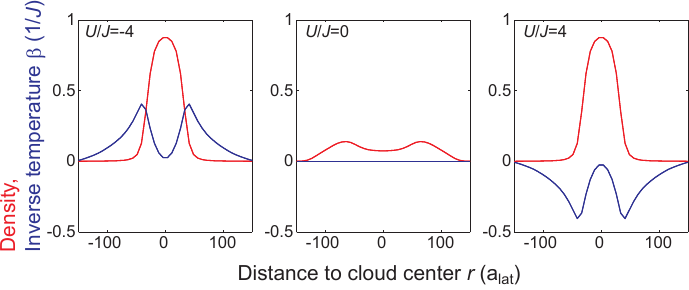}
\caption{Inverse temperature (blue) and density (red) distribution for a diagonal 
cut through the cloud after an expansion of $\tau=50\,\hbar/J$ for various interactions.
}\label{fig:NegT}
\end{figure}

As is shown in figure~\ref{fig:NegT}, the local inverse temperature $\beta=1/T$ depends 
strongly on the interaction strength: It vanishes in the non-interacting case 
$(\beta=0,\;T=\infty$) and also shows the 
expected $U\leftrightarrow-U$ symmetry: $\beta_{+U}=-\beta_{-U}<0$.\\
Assuming local thermalization, the observed $U\leftrightarrow-U$
  symmetry (see Fig.\ 3\&4 in main text) directly implies negative temperatures for repulsive
  interactions at long expansion times.

\subsection{Dynamical $U\leftrightarrow-U$ symmetry of the Hubbard model}
The observed $U\leftrightarrow-U$ symmetry arises from the
high symmetry of the tight binding dispersion $E_{\bf q} =
-2J \sum_i \cos(q_i \frac{\lambda}{2}) $ as can be understood by considering
 a single momentum component, say $\hbar q_x$, 
whose dispersion and group velocity distribution are plotted in Fig.\ 4b
in the main text.
The maximum group velocity $v_\text{max}=\text{max}(\frac{1}{\hbar}\frac{\partial E_q}{\partial q_x})$ 
corresponds to the state with quasi-momentum $\hbar q_x=1/2\,\hbar \pi/a_{lat}$ and
energy $E_q=0$, in the middle between the center and the edge of the Brillouin zone.
For Bloch waves with finite energy, either positive or
negative, the group velocity decreases symmetrically around this point.  
Intuitively speaking, the observed symmetry with respect to the change 
of the sign of $U$ can be understood as follows: 
Attractive interactions reduce the kinetic energy of the particles, which slows
them down. Repulsive interactions increase their kinetic energy which,
surprisingly, slows them down as well.

In the following we will turn this qualitative argument into a
rigorous theorem: 

We consider a coherent dynamical evolution 
arising from two Hubbard-type Hamiltonians that differ only in the sign of the
interaction term:
\begin{eqnarray}
{\hat{\cal H}}_{\pm} = -J \sum_{\langle ij \rangle\sigma } \hat{c}_{i\sigma}^\dagger \hat{c}_{j\sigma}  \pm U\sum_i \hat{n}_{i\uparrow}
\hat{n}_{i\downarrow} 
\end{eqnarray}
In addition, we define two symmetry operations: The time reversal operator 
$\hat{R}_t$ and  the $\pi$-boost operator $\hat{B}_Q$, which translates all quasi-momenta
by $\hbar Q=\hbar(\pi, \pi,\pi)\times\frac{1}{a_{lat}}$. Applying the time reversal operator turns the wavefunction into 
its complex conjugate~\cite{kim} or, equivalently, modifies the time evolution operator:
\begin{eqnarray}
\hat{R}_t e^{-i{\hat{\cal H }} t} \hat{R}_t^{\dagger} = e^{i{\hat{\cal H }} t}
\end{eqnarray}
The action of the $\pi$-boost operator is given in second quantized notations in quasi-momentum space by
\begin{eqnarray}
\hat{B}_Q \hat{c}_p \hat{B}_Q = \hat{c}_{p+Q}
\end{eqnarray}
or, in real space:
\begin{eqnarray}
\hat{B}_Q \hat{c}_j \hat{B}_Q = e^{iQr_j} \hat{c}_{j}
\end{eqnarray}
The Boost operator assigns an additional position-dependent phase $e^{iQr}$ to every Wannier state.

We now formulate the following general theorem:

\vspace*{\baselineskip}
\textit{If the experimentally measured quantity $\hat{O}$ is invariant under both time reversal and  $\pi$-boost, 
and the initial state $|\Psi_0\rangle$ is time reversal invariant and only acquires a global phase factor under 
the boost transformation $(\hat{B}_Q\ket{\Psi_0}=e^{i\chi}\ket{\Psi_0},\,\chi\in\mathbb{R})$, 
then the observed time evolutions} 
\begin{eqnarray}
\langle \hat{O}(t) \rangle_{\pm}  = 
\langle \Psi_0 | e^{i {\cal \hat{H}}_\pm t} \hat{O} e^{-i {\cal \hat{H}}_\pm t} | \Psi_0 \rangle
\end{eqnarray} 
 \textit{are identical: $\langle \hat{O}(t) \rangle_{+}= \langle \hat{O}(t)\rangle_{-}$.}

\vspace*{\baselineskip}
In order to prove the above symmetry theorem we first observe that 
\begin{eqnarray}
\langle \hat{O} (t) \rangle_+ &=& 
\langle \Psi_0 | \hat{R}_t^\dagger \hat{R}_t e^{i {\cal \hat{H}}_+ t} \hat{R}_t^\dagger \hat{R}_t \hat{O} 
\hat{R}_t^\dagger \hat{R}_t e^{-i {\cal \hat{H}}_+ t} \hat{R}_t^\dagger \hat{R}_t | \Psi_0 \rangle
\nonumber\\
&=& \langle \Psi_0 | e^{-i {\cal \hat{H}}_+ t} \hat{O} e^{i {\cal \hat{H}}_+ t} | \Psi_0 \rangle
\label{Tinvariance}
\end{eqnarray}
The last equation follows from the definition of time reversal invariance, $\hat{R}_t |\Psi_0\rangle = |\Psi_0\rangle$ and
$\hat{R}_t \hat{O} \hat{R}_t^{\dagger} = \hat{O}$, and from the unitarity property $\hat{R}_t^\dagger \hat{R}_t =1$.
Note that equation (\ref{Tinvariance}) corresponds to the symmetry of time evolutions for 
${\cal \hat{H}} \rightarrow -{\cal \hat{H}}$, which has been discussed previously in References~\cite{sorensen,cirac}.

From the definition of the $\pi$-boost we get:
\begin{eqnarray}
\hat{B}_Q {\cal \hat{H}}_{\pm} \hat{B}_Q &=&-J \sum_{\langle ij \rangle\sigma } \hat{B}_Q 
\hat{c}_{i\sigma}^\dagger \hat{B}_Q^2 \hat{c}_{j\sigma} \hat{B}_Q \\
&\quad&\pm U\sum_i \hat{B}_Q \hat{n}_{i\uparrow} \hat{B}_Q^2 \hat{n}_{i\downarrow} \hat{B}_Q\nonumber\\
&=& +J \sum_{\langle ij \rangle\sigma } \hat{c}_{i\sigma}^\dagger \hat{c}_{j\sigma}  
\pm U\sum_i \hat{n}_{i\uparrow} \hat{n}_{i\downarrow} \nonumber\\
&=& - {\cal \hat{H}}_{\mp}
\end{eqnarray}
Here we used the unitarity of the boost operator $\hat{B}_Q^2=1$ and the transformation behavior 
of the density operator $\hat{B}_Q \hat{n}_{i\updownarrow} \hat{B}_Q=\hat{B}_Q \hat{c}_{i\updownarrow}^\dagger 
\hat{B}_Q^2 \hat{c}_{i\updownarrow} \hat{B}_Q=\hat{n}_{i\updownarrow}$. 
With this we can continue equation (\ref{Tinvariance}):
\begin{align}
\langle \hat{O} (t)\rangle_+ &= 
\langle \Psi_0 | \hat{B}_Q^2  e^{-i {\cal \hat{H}}_+ t} \hat{B}_Q^2  \hat{O} 
\hat{B}_Q^2  e^{i {\cal \hat{H}}_+ t} \hat{B}_Q^2 | \Psi_0 \rangle\nonumber\\
&=\langle \Psi_0 | \hat{B}_Q  e^{+i {\cal \hat{H}}_- t}  \hat{O}  e^{-i {\cal 
\hat{H}}_- t} \hat{B}_Q | \Psi_0 \rangle\nonumber\\
&=\langle \Psi_0 | e^{-i\chi}  e^{+i {\cal \hat{H}}_- t}  \hat{O}  e^{-i {\cal 
\hat{H}}_- t}  e^{i\chi} | \Psi_0 \rangle\nonumber\\
&= \langle \hat{O} (t)\rangle_- e^{-i\chi}e^{i\chi}\nonumber\\
&= \langle \hat{O} (t)\rangle_-
\end{align}

In the last equation we used the $\pi$-boost invariance of the observable $\hat{B}_Q \hat{O} 
\hat{B}_Q = \hat{O}$, the required transformation behavior of the initial state 
$\hat{B}_Q\ket{\Psi_0}=e^{i\chi}\ket{\Psi_0}$, and the unitarity of the boost operator $\hat{B}_Q^2=1$. \hfill$\square$

\vspace*{\baselineskip}
The initial state given in the experiment can be written as an incoherent mixture of states of the form:
\begin{align}
\ket{\Psi_{mb}}=\prod_{i=1}^n \hat{c}_{r_i}^\dagger\ket{vac}\label{eqn:psimb}
\end{align}
This state describes $n$ particles localized at the positions $r_i$ and transforms under $\hat{B}_Q$ according to:
\begin{align}
\hat{B}_Q\ket{\Psi_{mb}}&=\hat{B}_Q\prod_{i=1}^n \hat{c}_{r_i}^\dagger\,\ket{vac}\nonumber\\
	&=\hat{B}_Q\, \hat{c}_{r_1}^\dagger \,\hat{B}_Q^2 \,\hat{c}_{r_2}^\dagger \,\hat{B}_Q^2\, 
	\hat{c}_{r_3}^\dagger\, \hat{B}_Q^2\cdots\, \hat{c}_{r_n}^\dagger \,\hat{B}_Q^2\,\ket{vac}\nonumber\\
	&=e^{iQr_1}\,\hat{c}_{r_1}^\dagger \,e^{iQr_2}\,\hat{c}_{r_2}^\dagger\, e^{iQr_3}\,
	\hat{c}_{r_3}^\dagger \cdots e^{iQr_n}\,\hat{c}_{r_n}^\dagger\,\ket{vac}\nonumber\\
	&=\prod_i e^{iQr_i}\ket{\Psi_{mb}}\nonumber\\
	&=\ e^{iQ\sum_i r_i}\ket{\Psi_{mb}}
	\end{align}
On the second line we used $\hat{B}_Q^2=1$ and on the third line we used $\hat{B}_Q\ket{vac}=\ket{vac}$.

An according calculation with $\hat{R}_t$ results in $\hat{R}_t\ket{\Psi}=\ket{\Psi}$.
This shows that a many-body state of the form of eqn.\ \ref{eqn:psimb} fulfills the requirements of the above symmetry theorem.
The extension to the mixed state used in the experiment is straightforward. From the definition of a general density matrix 
$\rho=\sum_j p_j \ket{\Psi_j}\bra{\Psi_j}$ we get:
\begin{align} 
\langle \hat{O}\rangle_\rho=tr[\rho \hat{O}]=\sum_j p_j \bra{\Psi_j}\hat{O}\ket{\Psi_j}
\end{align}
and see that the theorem also holds for mixed states, as it holds for every term in the sum.

The experimental observable is the density distribution $\hat{n}(r_j)=\sum_\sigma 
\hat{c}_{j\sigma}^\dagger  \hat{c}_{j\sigma}$ and the initial 
state consists of atoms that are completely localized to individual lattice sites (cf.\ sec.\ \ref{experimentalSequence}). 
Because both the initial state and the measured operator fulfill the requirements of the symmetry theorem, we are guaranteed to find 
the described $U \leftrightarrow -U$ symmetry in the dynamics for all interaction strengths. 

Since the bi-partite character of the lattice is crucial to the
proof of the theorem, this symmetry can be expected to be broken in lattices 
without the bi-partite structure, such as a triangular lattice.

\subsection{Image processing and fitting}\label{sec:fitting}
\subsubsection{Non-interacting case}
In the non-interacting case the perpendicular cloud size $\sqrt{\langle r^2\rangle}$ 
is extracted from in-situ phase-contrast images  using a 2D Gaussian fit:
\begin{equation}
G(x,y)=A\,e^{-\frac{(x-x_c)^2}{2\sigma_x^2}-\frac{(y-y_c)^2}{2\sigma_y^2}} +b
\end{equation}
Here $x_c,y_c,\sigma_x,\sigma_y,A,b$ are free fit parameters and the perpendicular 
cloud size is given by $\sqrt{\langle r^2\rangle}=R_G=\sqrt{\sigma_x^2+\sigma_y^2-w^2}$, 
where $w$ denotes the imaging resolution (radius of Airy disc $w< 3\mu$m) of our imaging setup.

The corresponding single-particle width $R_{sp}(t)$ is calculated by deconvolving
$R_G(t)$ with the initial width $R_G(0)$: $R_{sp}(t)=\sqrt{R_G(t)^2-R_G(0)^2}$. An example of the expanded
single-particle width is shown in Fig. 2 in the main text and agrees well with the linear slope expected for a
ballistic expansion.

In order to extract the \textit{mean expansion velocity} $v_\text{exp}$ the cloud sizes $R_G$ are fitted by:
\begin{equation}
R(t)=\sqrt{R_0^2+v_\text{exp}^2 t^2}\label{eqn:fitfct}
\end{equation} 

\subsubsection{Interacting case}

\begin{figure}[hbt]
\centering
\includegraphics[width=.9\columnwidth]{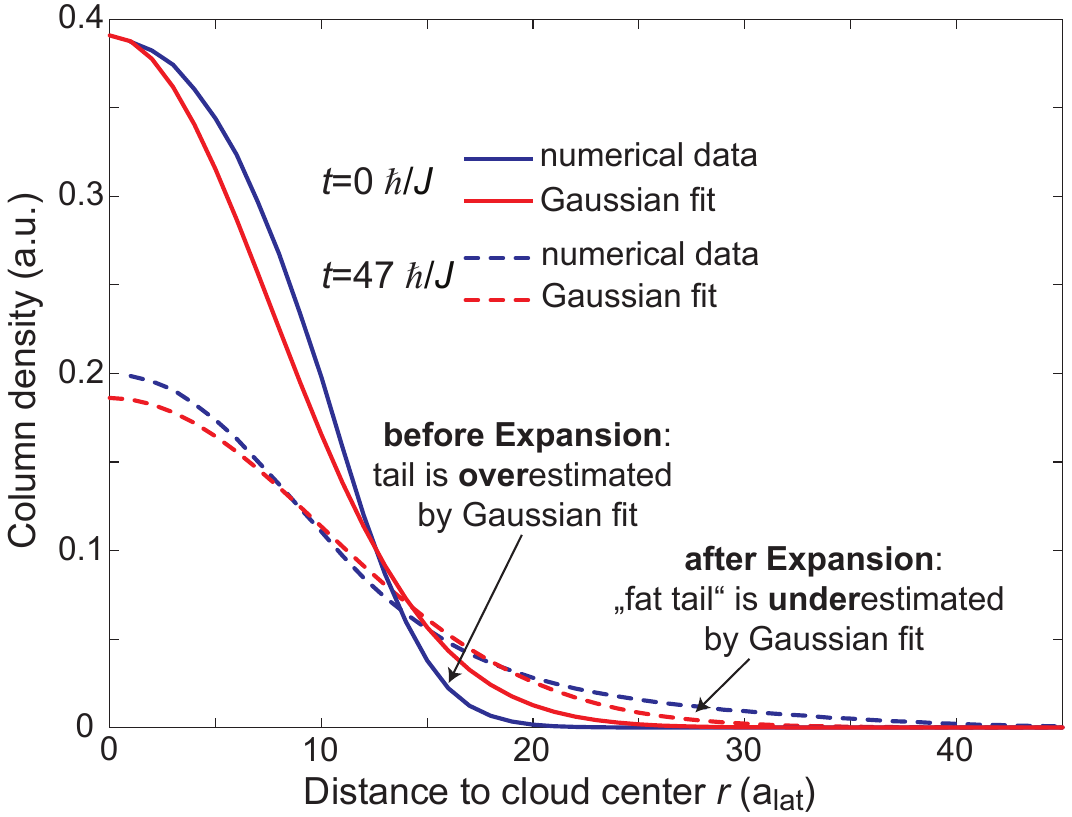}
\caption{Numerically calculated density distribution for $U/t=1.2$ together with Gaussian fits.}\label{fig:GF}
\end{figure}

In the case of interacting atoms the shape of the cloud changes considerably during the expansion, evolving from
a ``compact" Fermi-Dirac like shape (Supporting Online Material of Ref.~\cite{schneider:MI}) to a ``fat tail"
distribution, as illustrated in Fig.\ \ref{fig:GF} using fits to numerically simulated data. This leads to
considerable systematic errors in the estimation of $\langle r^2\rangle\,(\neq R_G^2)$ in the interacting case.
In principle, such systematic errors could be avoided by determining $\left\langle r^2\right\rangle$ via direct integration, but 
in the experiment this is hindered by imaging aberrations and the small signal to noise ratio in the extreme dilute limit.

The above change in the shape of the cloud is due to the density dependent dynamics in the interacting case: While
the expansion remains ballistic in the low density limit, where the mean free path is larger than the distance
to the cloud edge, the expansion velocity decreases for higher densities due to the increasing number of
collisions. As a consequence,  $\langle r^2\rangle$ will be dominated by the ballistically expanding outermost 
atoms for long expansion times.

\begin{figure}[hbt]
\centering
\includegraphics[width=1\columnwidth]{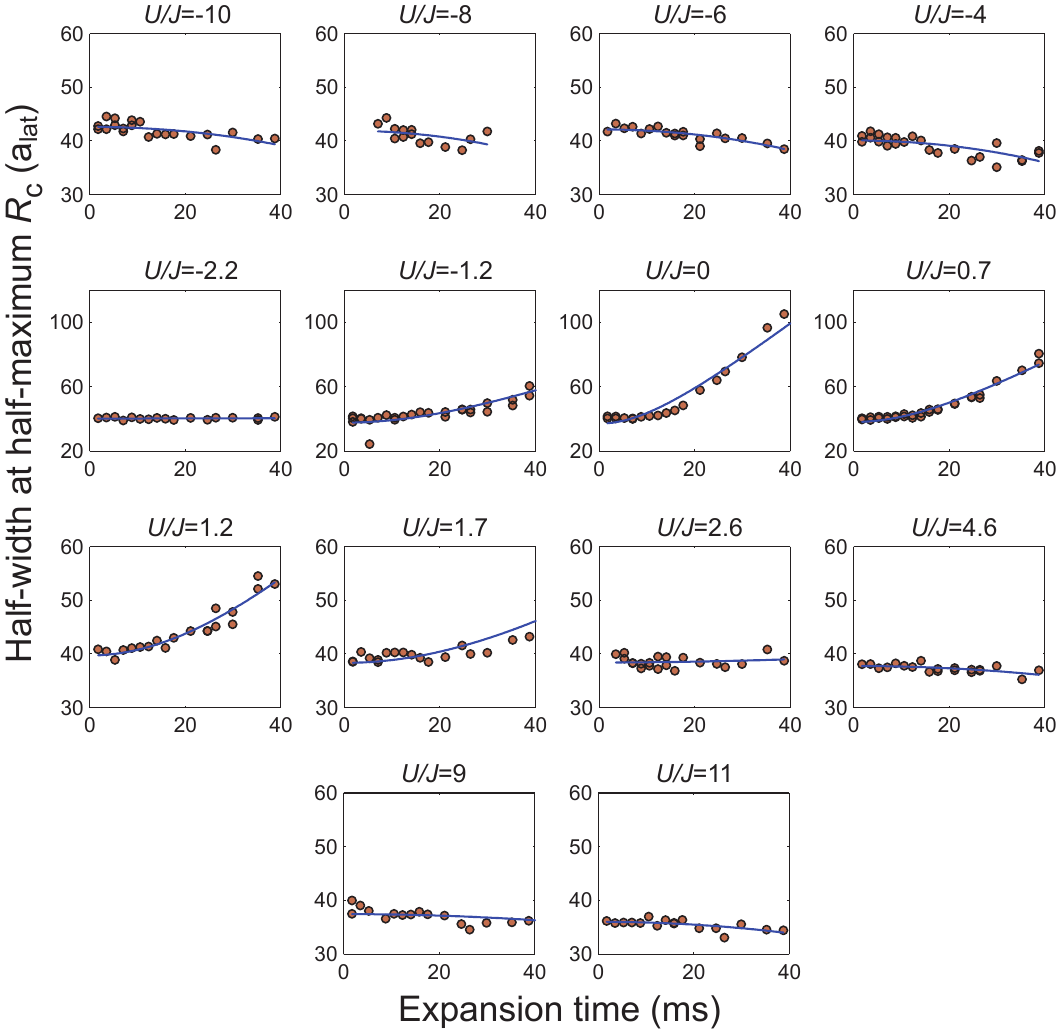}
\caption{Core widths $R_c$ as a function of expansion time for various interactions in an $8\,E_r$ 
deep lattice in the 2D case. Blue lines denote the fits used to extract the core expansion velocities 
(Eq.~\ref{eqn:fit})}\label{fig:HWHM}
\end{figure}

In order to focus on the core dynamics we instead use the \textit{core width} $R_c$, which denotes the half width at half maximum (HWHM)
 of the azimuthally averaged column density distribution, where each distribution is individually normalized.
These core widths are shown in Fig.\ \ref{fig:HWHM} and are fitted by the same fit function as in the non-interacting case 
\begin{equation}
R_c(t)=\sqrt{R_c(0)^2+v_c^2t^2}\label{eqn:fit}
\end{equation}
and the resulting core expansion velocities $v_c$ are shown in Fig.\ 5 in the main text.

\subsection{Expansion velocities of non-interacting atoms in 2D}
In the absence of collisions and additional potentials the Hubbard Hamiltonian 
consists only of the hopping term, which describes the tunneling of a
particle from one lattice site to a neighboring site with a rate $J/\hbar$ ($\sigma$: spin):
\begin{equation}
 {\hat{\cal H}}=-J\sum_{\left\langle i,j\right\rangle ,\sigma}\hat{c}_{i,\sigma}^\dagger \hat{c}_{j,\sigma}\label{eq:Tunneling}
 \end{equation}

\begin{figure}[hbt]
\centering
\includegraphics[width=\columnwidth]{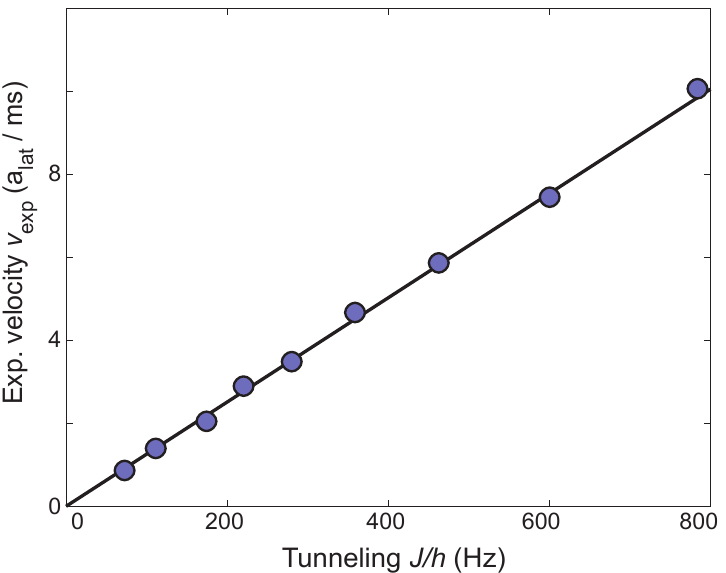}
\caption{Mean expansion velocity for different tunneling
  $J$. The solid line shows the quantum-mechanical prediction for 2D: $v_\text{exp}=2\,\frac{J}{\hbar}a_{lat}$. Statistical fit errors are comparable to symbol size.}\label{fig:CloudSize}
\end{figure}

The initially localized atoms expand ballistically with a mean squared velocity
$v_\text{exp}^2=\langle {\bf v}_{\bf q}^2 \rangle$, which is given by the mean
squared group velocity ${\bf v}_{\bf q}= \frac{1}{\hbar} \frac{\partial E}{\partial {\bf q}}$ of all Bloch waves.
Here $\hbar {\bf q}$ denotes the quasi-momentum
and $E({\bf q})=-2 J \sum_i \cos (q_i a_{lat})$ the dispersion relation in the lowest band. 
For particles that are initially localized to single lattice sites, the mean expansion velocity is
$v_\text{exp}=\sqrt{2d}\,\frac{J}{\hbar}a_{lat}$
($d$: dimension).  
This quantum mechanical prediction agrees very well with the extracted mean expansion velocity (see Fig.\ \ref{fig:CloudSize}) for various lattice depths, thereby verifying that the tunneling rate $J$ is the only energy scale left in the problem.

\subsection{Canceling the harmonic confinement}\label{sec:cancelHarm}
\begin{figure}[htb]
\centering
\includegraphics[width=\columnwidth]{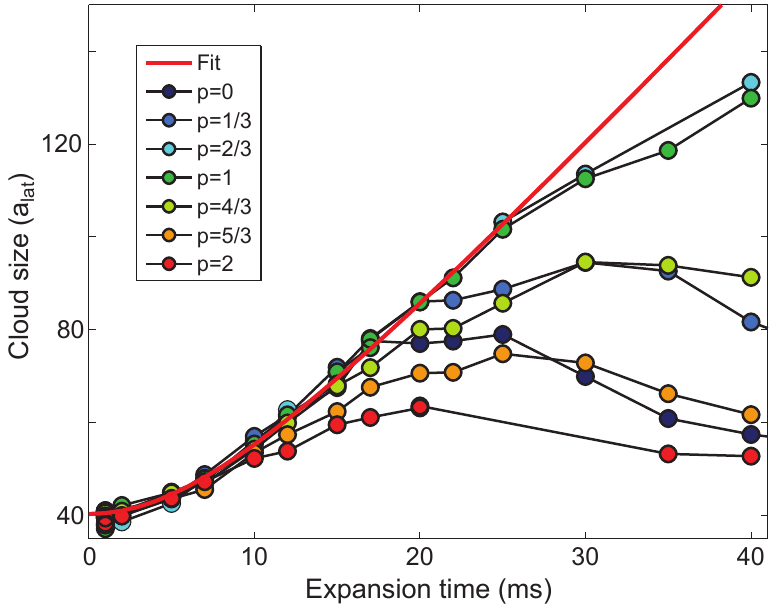}
\caption{Expansion of non-interacting atoms as a function of dipole beam power $p$ (a.u.) in an $8E_r$ lattice.
}\label{fig:NonIADipole}
\end{figure}

Figure~\ref{fig:NonIADipole} shows the cloud sizes $R_G(t)$ of an expanding non-interacting cloud in an
$8\,E_r$ deep quasi 2D lattice as a function of the dipole laser power during the expansion. 
The red line denotes a fit with the expected  dynamics (Eq.~\ref{eqn:fitfct}) to the first $20\,\text{ms}$. 
While the initial expansion velocity depends only slightly on the residual confinement, 
it completely dominates the size after long expansion times.

The largest cloud sizes are reached if the confinement created by the dipole trap 
precisely compensates the anticonfinement due to the lattice. 
This situation corresponds to dipole powers between  $p=2/3$ and $p=1$ in Fig.\ \ref{fig:NonIADipole}. 
Both an over- and an under-compensation lead to deviations from the expected ballistic
behavior and limit the cloud size by either classical reflections or
Bragg reflections of the expanding atoms. 

\begin{figure}[hbt]
\centering
\includegraphics[width=.8\columnwidth]{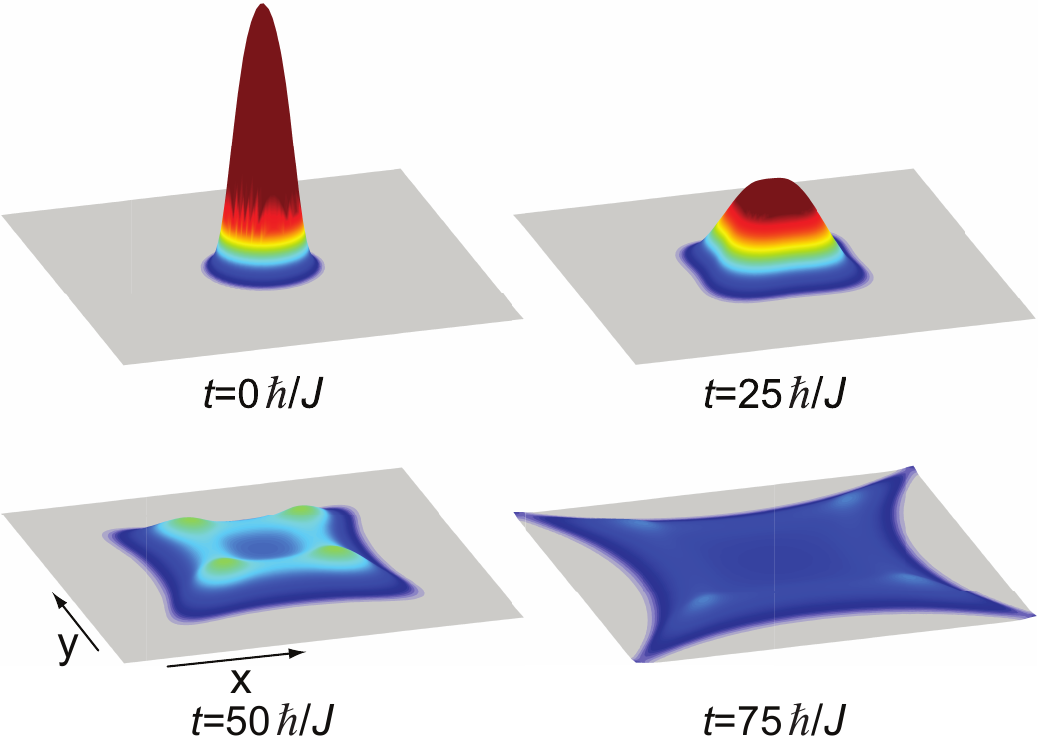}
\caption{Simulated column density distribution for the non-interacting expansion including the finite size 
of the lattice beams. }\label{fig:Jvar}
\end{figure}

In the well-compensated case the leading correction to the homogeneous Hubbard model
 arises from the fact that the hopping rate in any given
direction becomes larger when an atom is not in the center of the 
corresponding laser beam any more.  This effect can
easily be described quantitatively by modeling the experimental laser
profiles. For a distance of 100 lattice constants from the center, the
hopping rate increases by about 25\%. Figure~\ref{fig:Jvar} shows the resulting
density profiles in the non-interacting case,
which reproduce all features of Fig.\ 2 of the main text.

In the interacting case the  clouds remain much smaller
and these effects can completely be neglected.

In the $2D$ case the vertical lattice is kept at a lattice depth of $V_v=20\,E_r$, 
which corresponds to a (resonant) tunneling rate of $J_v=h\times 22\,\text{Hz}$.
While the resulting timescale $\tau=\hbar/J_v\approx7\,\text{ms}$ for resonant tunneling is shorter then the
used expansion times, gravity provides a strong potential in the vertical direction and leads to Bloch oscillation 
whose amplitude $2J/mg\ll\lambda/2$ is much smaller than the lattice constant, thereby isolating the individual 2D systems.

\subsection{Width of Feshbach resonance}\label{sec:FR}
The interaction strength during the expansion is controlled by use of the well-known Feshbach resonance at 
$B_0=202.1\,\text{G}$ between the two lowest hyperfine states $|\smash F,m_F\rangle=|9/2,-9/2\rangle$ and
$|9/2,-7/2\rangle$~\cite{JILA:FR}.
Compared to previous dipole trap experiments, the dynamics in the lattice are much more sensitive
to small scattering lengths, since the strongly reduced kinetic energy of atoms in a 
lattice enhances the role of interactions.
The observed pronounced dependence on small interactions (Fig.\ 4 in main text) 
enabled us to remeasure the zero crossing of the scattering length, which we found
to be at $B(a=0)=209.1\pm0.2\,\text{G}$.
Using the standard parametrization of the (free space) Feshbach resonances
\begin{equation}
a(B)=a_{\text{bg}}\left( 1-\dfrac{w}{B-B_0}\right)
\end{equation}
this zero crossing leads to a new width of
\begin{equation}
w=7.0\pm0.2\,\text{G}
\end{equation}
 compared to the previous dipole trap measurement
of $w_\text{dipole}=7.8\pm0.6\, \text{G}$~\cite{Regal}. In addition to the pronounced dependence of the slope, 
only the expansion at the newly assigned zero crossing matches that of a single component 
Fermi gas under the same conditions and leads to the square shape expected for non-interacting atoms.
The remaining uncertainty in the position of the zero crossing is dominated by uncertainties in the magnetic field calibration.

\begin{figure}[ht]
\centering
\includegraphics[width=\columnwidth]{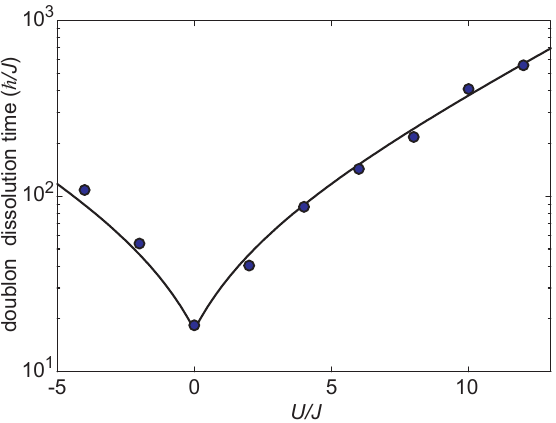}
\caption{Measured doublon dissolution times. The black line is an exponential fit and serves as guide to the eye.}\label{fig:DD}
\end{figure}

\subsection{Doublon dissolution time}
Both for the qualitative and quantitative analysis of our results, we
assume the relaxation of the system to local equilibrium.
For very strong attractive or repulsive interactions $|U|\gg J$, however, 
doubly occupied sites (doublons) only decay very slowly~\cite{Winkler:RBP,Rosch:Metastable,eth:doublondecay} 
as the missing or excess energy of order $U$ cannot easily be transferred 
to other particles.  

An important question is whether the rate with which the diffusive core 
melts is determined by the decay time of 
\textit{individual} doublons or whether the latter is fast compared to the former.
The number of atoms on doubly occupied lattice sites is measured by converting all doublons into
molecules. This is achieved by first increasing the lattice depth to $20\,E_r$ in $200\,\mu\text{s}$,
followed by a magnetic field ramp ($5\,\text{G/ms}$) over the Feshbach resonance~\cite{schneider:MI}. 
The resulting doublon numbers are fitted by a simple exponential decay.
The resulting doublon dissolution times (Fig.\ \ref{fig:DD}) 
are about an order of magnitude larger
than the decay time of excess doublons measured recently in a half
filled situation~\cite{eth:doublondecay} in 3D at comparable
interactions. Furthermore, they match very well the typical timescales for melting 
of the diffusive core of the cloud observed within our numerical
simulations. We therefore conclude that for the parameters
used in our experiment the doubly occupied sites remain in local
equilibrium in the diffusive regime.

\section{Acknowledgments}
We thank Maria Moreno-Cardoner, Fabian Heidrich-Meisner, 
David Pekker and Rajdeep Sensarma, Bernd Kawohl,
Claus Kiefer, Joachim Krug and Martin Zirnbauer for
stimulating and insightful discussions.

This work was supported by the Deutsche Forschungsgemeinschaft 
(FOR801, SFB TR 12, SFB 608, Gottfried Wilhelm Leibniz Prize), 
the European Union (Integrated Project SCALA), EuroQUAM (L.H.), 
the US Defense Advanced Research Projects Agency 
(Optical Lattice Emulator program), 
the US Air Force Office of Scientific Research 
(Quantum Simulation MURI (E.D.)), 
the National Science Foundation (DMR-07-05472) (E.D.), 
the Harvard-MIT CUA (E.D.), MATCOR (S.W.), the Gutenberg Akademie (S.W.)
and the German National Academic Foundation (S.M.).

\section{Author contribution}
U.S., L.H. and J.P.R. carried out the measurements, U.S. performed the data analysis with contributions from L.H. and J.P.R..
I.B. supervised the measurements. S.M. and D.R. performed the numerical calculations supervised by A.R.. E.D., U.S. and A.R. constructed the analytical proof of the dynamical symmetry.
U.S. and A.R. wrote the manuscript with substantial contributions by I.B. and all authors.

\section{Additional Information}
The authors declare that they have no competing financial interests. Correspondence and requests for materials should be addressed to U.S.~(email: ulrich.schneider@lmu.de)


\begin{thebibliography}{10}

\bibitem{Jaksch}
Jaksch,~D. \& Zoller, P. The cold atom Hubbard toolbox.
{\em Ann. Phys.}, \textbf{315,} 52--79 (2005).

\bibitem{Lewenstein}
Lewenstein,~M. {\em et al.} Ultracold atomic gases in optical lattices: mimicking condensed
  matter physics and beyond. {\em Adv. Phys.}, \textbf{56,} 243--379 (2007)

\bibitem{Bloch:RMP}
Bloch,~I., Dalibard,~J. \& Zwerger,~W. Many-body physics with ultracold gases.
{\em Rev. Mod. Phys.}, \textbf{80,} 885 (2008)

\bibitem{Hung2010}
Hung,~C., Zhang,~X., Gemelke,~N. \& Chin,~C. 
Slow Mass Transport and Statistical Evolution of an Atomic Gas
  across the Superfluid-Mott-Insulator Transition.
{\em Phys. Rev. Lett.}, \textbf{104,} 160403 (2010)

\bibitem{eth:doublondecay}
Strohmaier,~N. {\em et al.} Observation of elastic doublon decay in the Fermi-Hubbard model.
{\em Phys. Rev. Lett.}, \textbf{104,} 080401 (2010)

\bibitem{Julia}
Wernsdorfer,~J., Snoek,~M. \& Hofstetter,~W. 
Lattice-ramp-induced dynamics in an interacting bose-bose mixture.
{\em Phys. Rev. A}, \textbf{81,} 043620 (2010)

\bibitem{Tatjana}
Gericke,~T.  {\em et al.} 
Adiabatic loading of a Bose-Einstein condensate in a 3D optical
  lattice.
{\em J. Mod. Opt.}, \textbf{54,} 735 (2007)

\bibitem{Hackermueller}
Hackerm\"{u}ller,~L. {\em et al.}  
Anomalous Expansion of Attractively Interacting Fermionic Atoms in
  an Optical Lattice.
{\em Science}, \textbf{327,} 1621 (2010)

\bibitem{ETHZ:MI}
J\"{o}rdens,~R., Strohmaier,~N., G\"{u}nter,~K., Moritz,~H. \& Esslinger,~T. 
A Mott insulator of fermionic atoms in an optical lattice.
{\em Nature}, \textbf{455,} 204 (2008)

\bibitem{schneider:MI}
Schneider,~U.  {\em et al.} 
Metallic and Insulating Phases of Repulsively Interacting Fermions
  in a 3D Optical Lattice.
{\em Science}, \textbf{322,} 1520 (2008)

\bibitem{Hubbard}
Hubbard,~J. 
Electron Correlations in Narrow Energy Bands.
{\em Proc. R. Soc. A}, \textbf{276,} 238 (1963)

\bibitem{LENS:insul}
Pezz\`{e},~L.,   {\em et al.} 
Insulating Behavior of a Trapped Ideal Fermi Gas.
{\em Phys. Rev. Lett.}, \textbf{93,} 120401 (2004)

\bibitem{LENS:collInduced}
Ott,~H., {\em et al.}  
Collisionally Induced Transport in Periodic Potentials.
{\em Phys. Rev. Lett.}, \textbf{92,} 160601 (2004)

\bibitem{ETHZ:transport}
Strohmaier,~N., {\em et al.} 
Interaction-Controlled Transport of an Ultracold Fermi Gas.
{\em Phys. Rev. Lett.}, \textbf{99,} 220601 (2007)

\bibitem{Lignier:dynCont}
Lignier,~H., {\em et al.}  
Dynamical Control of Matter-Wave Tunneling in Periodic Potentials.
{\em Phys. Rev. Lett.}, \textbf{99,} 220403 (2007)

\bibitem{Dahan:BlochOsc}
Ben~Dahan,~M., Peik,~E., Reichel,~J., Castin,~Y. \& Salomon,~C. 
Bloch Oscillations of Atoms in an Optical Potential.
{\em Phys. Rev. Lett.}, \textbf{76,} 4508 (1996)

\bibitem{porto:osc}
Fertig,~C.D., {\em et al.} 
Strongly Inhibited Transport of a Degenerate 1D Bose Gas in a
  Lattice.
{\em Phys. Rev. Lett.}, \textbf{94,} 120403 (2005)

\bibitem{gustavsson:BlochOsc}
Gustavsson,~M., {\em et al.}
Control of Interaction-Induced Dephasing of Bloch Oscillations.
{\em Phys. Rev. Lett.}, \textbf{100,} 080404 (2008)

\bibitem{fattori:BlochOsc}
Fattori,~M., {\em et al.} 
Atom Interferometry with a Weakly Interacting Bose-Einstein
  Condensate.
{\em Phys. Rev. Lett.}, \textbf{100,} 080405 (2008)


\bibitem{Aharonov}
Aharonov,~Y., Davidovich,~L. \& Zagury,~N. 
Quantum random walks.
{\em Phys. Rev. A}, \textbf{48,} 1687 (1993)

\bibitem{Farhi:ContQW}
Farhi,~E. \& Gutmann,~S. 
Quantum computation and decision trees.
{\em Phys. Rev. A}, \textbf{58,} 915 (1998)

\bibitem{widera:quantumwalk}
Karski,~M., {\em et al.} 
Quantum Walk in Position Space with Single Optically Trapped Atoms.
{\em Science}, \textbf{325,} 174 (2009)

\bibitem{Weitenberg:addres}
Weitenberg,~C.,  {\em et al.}  
Single-Spin Addressing in an Atomic Mott Insulator.
{\em Nature}, \textbf{471,} 319 (2011)


\bibitem{Childs:QWA}
Childs,~A.M., {\em et al.}   
Exponential algorithmic speedup by a quantum walk.
{\em STOC '03: Proceedings of the thirty-fifth annual ACM
  symposium on Theory of computing}, 59 (2003)


\bibitem{Rigol:2Dtherm}
Rigol,~M., Dunjko,~V. \& Olshanii,~M. 
Thermalization and its mechanism for generic isolated quantum
  systems.
{\em Nature}, \textbf{452,} 854 (2008)

\bibitem{werner:Therm}
Eckstein,~M., Kollar,~M. \& Werner,~P. 
Thermalization after an Interaction Quench in the Hubbard Model.
{\em Phys. Rev. Lett.}, \textbf{103,} 056403 (2009)

\bibitem{Mandt2011}
Mandt,~S., Rapp,~A. \& Rosch,~A. 
Interacting fermionic atoms in optical lattices diffuse symmetrically upwards and
downwards in a gravitational potential.
{\em Phys. Rev. Lett.}, \textbf{106,} 250602 (2011)

\bibitem{vasquez06}
V\'azquez,~J.L. 
\textit{Smoothing and Decay Estimates for Nonlinear Diffusion
  Equations.}
Oxford University Press, Oxford (2006)

\bibitem{Sanpera:ent}
Romero-Isart,~O., Eckert,~K., Rodo,~C. \& Sanpera,~A. 
Transport and entanglement generation in the Bose-Hubbard model.
{\em J. Phys. A}, \textbf{40,} 8019 (2007)

\bibitem{Fabian:QDist}
Heidrich-Meisner,~F., {\em et al.}
Quantum distillation: Dynamical generation of low-entropy states of
  strongly correlated fermions in an optical lattice.
{\em Phys. Rev. A}, \textbf{80,} 041603 (2009)

\bibitem{Schollwoeck2005}
Schollw\"ock,~U. 
The density.matrix renormalization group.
{\em Rev. Mod. Phys.}, \textbf{77,} 259 (2005)

\bibitem{Kajala2011}
Kajala,~J., Massel,~J. \& T\"orm\"a,~P. 
Expansion Dynamics in the One-Dimensional Fermi-Hubbard Model.
{\em Phys. Rev. Lett.}, \textbf{106,} 206401 (2011)

\bibitem{purcell51}
Purcell, E.M. \& Pound,~R.V. 
A nuclear spin system at negative temperature.
{\em Phys. Rev.}, \textbf{81,} 279 (1951)

\bibitem{Medley:11}
Medley,~P., Weld,~D.M., Miyake,~H., Pritchard,~D.E. \& Ketterle,~W. 
Spin Gradient Demagnetization Cooling of Ultracold Atoms.
{\em Phys. Rev. Lett.}, \textbf{106,} 195301 (2011)
  
\bibitem{Rapp2010}
Rapp,~A., Mandt,S. \& Rosch,~A. 
Equilibration Rates and Negative Absolute Temperatures for Ultracold
  Atoms in Optical Lattices.
{\em Phys. Rev. Lett.}, \textbf{105,} 220405 (2010)

\bibitem{McKay2010}
McKay,~D.C. \& DeMarco,~B. 
Cooling in strongly correlated optical lattices: prospects and
  challenges.
{\em Rep. Prog. Phys.}, \textbf{74,} 054401 (2011)

\bibitem{JILA:FR}
Loftus,~T., Regal,~C.A., Ticknor,~C., Bohn,~J.L. \& Jin,~D.S. 
Resonant Control of Elastic Collisions in an Optically Trapped Fermi Gas of Atoms.
{\em Phys. Rev. Lett.}, \textbf{88,} 173201 (2002)


\bibitem{will:multiorb}
Will,~S., \textit{et al.}
Time-resolved observation of coherent multi-body interactions in quantum phase revivals.
{\em Nature}, \textbf{465,} 197 (2010)

\bibitem{ziman}
Ziman,~J.M. 
\textit{Electrons and Phonons.}
Oxford University Press, New York (1960)

\bibitem{kim}
Kim,~S.K. 
\textit{Group theoretical methods}.
Cambridge University Press, Cambridge (1999)

\bibitem{sorensen}
Sorensen,~A.S.,  \textit{et al.}
Adiabatic preparation of many-body states in optical lattices.
{\em Phys. Rev. A}, \textbf{81,} 061603(A) (2010)

\bibitem{cirac}
Garc\'\i{}a-Ripoll,~J.J, Martin-Delgado,~M.A. \& Cirac,~J.I. 
Implementation of Spin Hamiltonians in Optical Lattices.
{\em Phys. Rev. Lett.}, \textbf{93,} 250405 (2004)

\bibitem{Regal}
Regal,~C. 
Experimental realization of BCS-BEC crossover physics with a Fermi gas of atoms.
PhD Thesis, University of Colorado, Boulder (2006)

\bibitem{Winkler:RBP}
Winkler,~K., \textit{et al.} 
Repulsively bound atom pairs in an optical lattice.
{\em Nature}, \textbf{441,} 853 (2006)

\bibitem{Rosch:Metastable}
Rosch,~A., Rasch,~D., Binz,~B. \& Vojta,~M. 
Metastable Superfluidity of Repulsive Fermionic Atoms in Optical
  Lattices.
{\em Phys. Rev. Lett.}, \textbf{101,} 265301 (2008)

\end{thebibliography}
\end{document}